\documentclass[aip, apl,%
reprint,%
]{revtex4-1}

\usepackage{graphicx}
\usepackage{dcolumn}
\usepackage{bm}

\begin{document}


\title{Mitigation of Cosmic Ray Effect on Microwave Kinetic Inductance Detector Arrays}

\author{K. Karatsu}
	\email{K.Karatsu@sron.nl}
	\affiliation{SRON Netherlands Institute for Space Research, Sorbonnelaan 2, 3584CA Utrecht, The Netherlands}
\author{A. Endo}
	\affiliation{Department of Microelectronics, Faculty of Electrical Engineering, Mathematics and Computer Science, Delft University of Technology, Mekelweg 4, 2628CD Delft, The Netherlands}
	\affiliation{Kavli Institute of Nanoscience, Faculty of Applied Sciences, Delft University of Technology, Lorentzweg 1, 2628CJ Delft, The Netherlands}
\author{J. Bueno}
	\affiliation{SRON Netherlands Institute for Space Research, Sorbonnelaan 2, 3584CA Utrecht, The Netherlands}
\author{P. J. de Visser}
	\affiliation{SRON Netherlands Institute for Space Research, Sorbonnelaan 2, 3584CA Utrecht, The Netherlands}
\author{R. Barends}
	\affiliation{Google, Santa Barbara, CA 93117, USA}
\author{D.J. Thoen}
	\affiliation{Department of Microelectronics, Faculty of Electrical Engineering, Mathematics and Computer Science, Delft University of Technology, Mekelweg 4, 2628CD Delft, The Netherlands}
	\affiliation{Kavli Institute of Nanoscience, Faculty of Applied Sciences, Delft University of Technology, Lorentzweg 1, 2628CJ Delft, The Netherlands}
\author{V. Murugesan}
	\affiliation{SRON Netherlands Institute for Space Research, Sorbonnelaan 2, 3584CA Utrecht, The Netherlands}
\author{N. Tomita}
	\affiliation{Department of Physics, School of Science, The University of Tokyo, 7-3-1 Hongo, Bunkyo-ku, Tokyo 113-0033, Japan}
\author{J.J.A. Baselmans}
	\affiliation{SRON Netherlands Institute for Space Research, Sorbonnelaan 2, 3584CA Utrecht, The Netherlands}
	\affiliation{Department of Microelectronics, Faculty of Electrical Engineering, Mathematics and Computer Science, Delft University of Technology, Mekelweg 4, 2628CD Delft, The Netherlands}

\date{\today}

\begin{abstract}
For space observatories, the glitches caused by high energy phonons created by the interaction of cosmic ray particles with the detector substrate lead to dead time during observation.
Mitigating the impact of cosmic rays is therefore an important requirement for detectors to be used in future space missions.
In order to investigate possible solutions, we carry out a systematic study by testing four large arrays of Microwave Kinetic Inductance Detectors (MKIDs), each consisting of $\sim$ 960 pixels and fabricated on monolithic 55~mm $\times$ 55~mm $\times$ 0.35~mm Si substrates.
We compare the response to cosmic ray interactions in our laboratory for different detector arrays: A standard array with only the MKID array as reference; an array with a low $T_c$ superconducting film as phonon absorber on the opposite side of the substrate; and arrays with MKIDs on membranes.
The idea is that the low $T_c$ layer down-converts the phonon energy to values below the pair breaking threshold of the MKIDs, and the membranes isolate the sensitive part of the MKIDs from phonons created in the substrate.
We find that the dead time can be reduced up to a factor of 40 when compared to the reference array.
Simulations show that the dead time can be reduced to below 1~\% for the tested detector arrays when operated in a spacecraft in an L2 or a similar far-Earth orbit.
The technique described here is also applicable and important for large superconducting qubit arrays for future quantum computers.
\end{abstract}

\pacs{Valid PACS appear here}
\maketitle

Data loss caused by cosmic ray hits on the instruments' detectors is one of the main concerns for space observatories.
Cosmic rays are so energetic that they penetrate the satellite structure, reach the detectors, and deposit a fraction of their energy through ionization and atomic excitation.
The deposited energy causes a cascade of high-energy phonons (ballistic phonons) that spread inside the detectors, trigger the detectors' response, and create glitches in the data stream.
These glitches lead to significant dead time of the detectors and loss of integration efficiency \cite{Planck2014}.
Although cosmic ray interactions are strongest  in space, even on earth the interaction rate is about a few events/min/cm$^2$, posing possible issues especially for cryogenic superconducting circuits with large areas, such as quantum computing chips.

Microwave Kinetic Inductance Detectors (MKIDs) \cite{Day2003, Zmuidzinas2012} are pair breaking detectors that sense a change in Cooper pair density due to radiation absorption, in contrast to bolometric detectors, such as Transition Edge Sensors \cite{Irwin1995}, which measure temperature.
Hence, MKIDs only sense energies larger than the gap energy of a superconductor film ($2 \Delta$).
In the Planck satellite, in which bolometers are used, short and long glitches are observed due to cosmic rays \cite{Catalano2014}.
Short glitches, with a decay of 4-10~ms, are determined to be direct hits in the bolometer.
Long glitches, with a time scale exceeding 10~ms, are interpreted as cosmic ray hits in the Si wafer.
The measurements of cosmic ray interactions on arrays of MKIDs on earth show interactions with two time scales: the ballistic phonon thermalization time, which is in the order of 0.1~ms \cite{Catalano2016}, and the quasiparticle recombination time ($\tau_{qp}$).
The latter depends on the background loading on the detector and varies between 0.01~ms for the high power background of a ground based imaging system to values in excess of 1~ms for low background applications in space \cite{deVisser2014}.
The dead time fraction caused by cosmic rays is therefore expected to be smaller in case of MKIDs.
\begin{figure*} 
\includegraphics{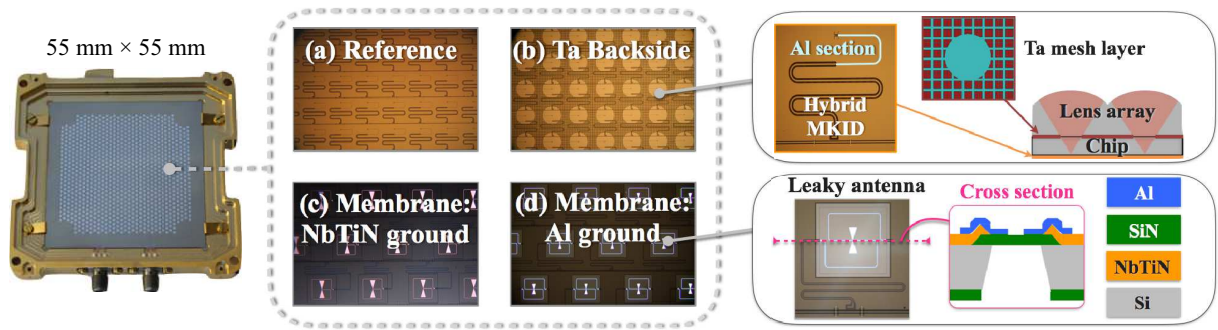}
\caption{\label{fig:1}
Images of measured MKIDs arrays.
(a) Reference array with classic design \cite{Janssen2013}.
(b) Array with Ta mesh layer.
The right top panel shows the relative position of lens array, MKID layer, and the mesh layer \cite{Baselmans2017, Yates2017}.
(c), (d) Arrays with membrane structures.
The right bottom panel shows the cross section of the membrane (not to scale) \cite{Hahnle2018, Bueno2017a, Bueno2017b}.
In chip $d$ the ground plane around the antenna is aluminium, for the rest it is made from NbTiN.
For those chips, no lens array was mounted during the measurement.
However, during nominal operation, a lens array is placed with a small gap above the antenna feed (see Ref.~\onlinecite{Bueno2017a})
}
\end{figure*}
Adding a superconducting layer with a critical temperature ($T_c$) below or close to the $T_c$ of the material of the MKID is an idea to further harden MKID arrays against the effects of cosmic ray hits by reducing the phonon energy \cite{Monfardini2016}.
In the low $T_c$ layer, the energies of high energy ballistic phonons are converted by electron-phonon interaction into phonons with energies mostly in between 2 to $3\Delta_{layer}$ \cite{Kozorezov2000, Goldie2013}.
Phonons emitted due to scattering will have even lower energy than $ 2\Delta_{layer}$.
Thus if the $T_c$ of the additional layer is chosen $T_c^{layer} \lesssim 2/3~T_c^{MKID}$, most of the cosmic ray energy will be quickly converted into low energy phonons to which the MKID is not sensitive.
Making the detector on a membrane structure is another idea to harden the MKID array for cosmic ray interactions by suppressing the propagation of high energy phonons towards the detector \cite{Holmes2008}.

In this letter, we measure the dead time due to cosmic ray interactions in four types of NbTiN-Al hybrid MKID arrays \cite{Janssen2013, Baselmans2017}.

All arrays are made on a 55~mm $\times$ 55~mm $\times$ 0.35~mm chip and contain about 960 hybrid MKIDs \cite{Janssen2013}, fabricated from NbTiN with a $T_c=15$~K, in which a small Al section of about 1.2 mm long ($T_c=1.25$~K) is used as the radiation detection area.
It needs to be mentioned that we sense only pair breaking events in this aluminium; pair breaking and subsequent recombination in the NbTiN area of the detector is much faster than the detector time constant and is thereby efficiently filtered.
Figure~\ref{fig:1} displays the 4 configurations of measured chips:
\begin{enumerate}
\setlength{\itemsep}{1pt}
\setlength{\parskip}{1pt}
\item[($a$)] A reference array with the NbTiN-Al MKIDs fabricated on a solid sapphire substrate without any backside layer or membrane structure \cite{Janssen2013}.
\item[($b$)] An array with a Ta ($T_c=0.65~{\rm K} \approx T_c^{Al}/2$) mesh layer on the backside \cite{Baselmans2017, Yates2017}.
\item[($c$)] An array with the Al section of the MKID on a membrane of 1 $\mu$m thick SiN (ground of the resonator is NbTiN).
Each MKID is coupled to a leaky lens antenna \cite{Hahnle2018}.
\item[($d$)] A membrane-based array similar to chip $c$: The difference is that the ground plane of the MKID on the membrane is also made from the same Al film \cite{Bueno2017a, Bueno2017b}.
 \end{enumerate}

\begin{figure}
\includegraphics{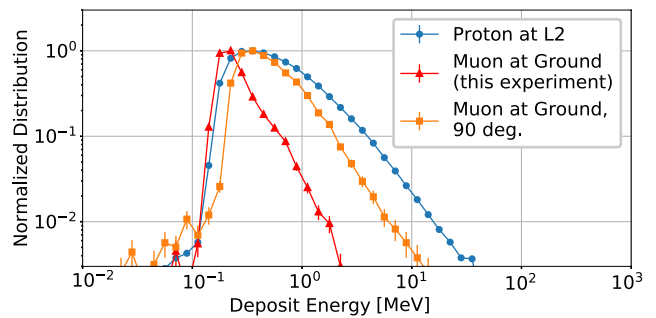}
\caption{\label{fig:2}
Simulated energy deposition by cosmic rays in an MKID chip (55~mm $\times$ 55~mm $\times$ 0.35~mm).
Blue, red, and orange lines show energy deposition by protons at an L2 orbit, muons at ground level with the chip placed horizontally, and muons at ground level with the chip placed vertically to increase interaction length, respectively.
}
\end{figure}

\begin{figure*} 
\includegraphics{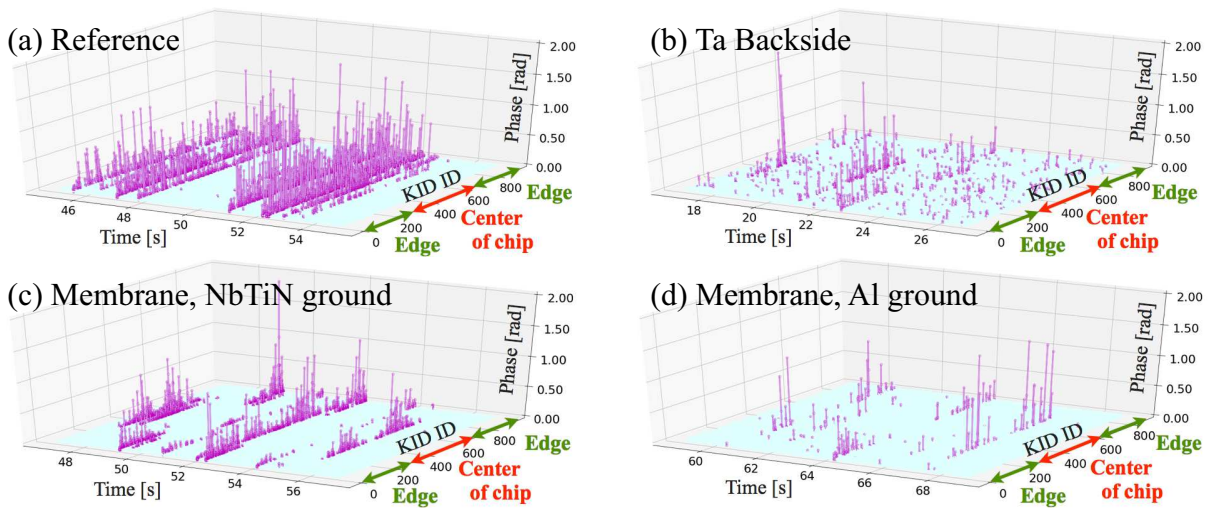}
\caption{\label{fig:3}
Typical 10~s time traces of phase response for different types of array.
Magenta points show the MKID response which exceeds the $5~\sigma$ threshold (glitches).
(a) Time trace of chip $a$.
A large fraction of the pixels are simultaneously affected by each cosmic ray event.
(b) Time trace of chip $b$.
While the spread of cosmic ray events in the chip is suppressed, quite many scattered events with small responses are observed that are poorly understood.
Those events only appear at a few pixels at the same time while normal cosmic ray events are always observed by more than 10 pixels simultaneously.
Detailed analysis confirm that these events are due to quasiparticle creation, as they are correlated in phase and amplitude.
Also, the events cannot be explained by statistics: Changing the threshold of glitch detection has a very small effect on the number count.
The overall effect of these events on the dead time is very small due to their isolated nature.
(c) Time trace of chip $c$.
The spread of cosmic ray events in the chip are suppressed.
(d) Time trace of chip $d$.
The effect of cosmic ray hits are highly suppressed, even when compared to chip $c$.
}
\end{figure*}

We evaluate the effect of cosmic ray interactions in the detector chips by measuring the effects of secondary cosmic rays in the laboratory.
The chips are placed horizontally in a dark, cold (120~mK) environment using an Adiabatic Demagnetization Refrigerator (ADR) with a box-in-box light-tight setup \cite{Baselmans2012}$^{,}$
\footnote{Chip $a$ and $b$ has a gold layer at edges to have a good thermal contact to the gold-plated sample box.
The layer is intended to better thermalize the chip under sub-mm illumination \cite{Yates2017}.}.
In order to accumulate statistics, we took 30 minutes ($T_{meas}$) of time ordered data for each chip with a multiplexing readout system \cite{Rantwijk2016} with a sampling rate of $\sim1.27$~kHz ($\Delta t_{read} = 786~\mu$s).
We can resolve the shape of the glitches with this sampling speed and evaluate $\tau_{qp}$ by fitting the glitch tails with a single exponential function \cite{Baselmans2017}.
Considering the typical muon rate of a few events/min/cm$^2$ at the ground, we expect about a few thousands cosmic ray events observed in each chip.
The main component of secondary cosmic rays is muons that result from the interaction of primary cosmic rays (mainly protons) with the Earth's atmosphere.
Taking into account the geometry of the cryostat, we simulate the energy deposition by cosmic rays in our detector chip by the GEANT4 simulation toolkit \cite{GEANT2003} with CRY database \cite{CRY}.
The simulated energy distribution is shown in Fig.~\ref{fig:2}.
We obtain a broad spectrum that peaks around 200 to 300~keV, which is far above the energy threshold of an Al MKID ($2 \Delta \sim 0.3$~meV).
The energy uncertainty of Fig.~\ref{fig:2} is about 10~\%. This mainly comes from the discrepancy between measured data and CRY database \cite{CRY}.
In order to identify cosmic ray events in the data, we use an iterative scheme based on the $2^{nd}$ derivative of the time trace data, by which glitches are enhanced as well as low frequency noise are removed.
The iteration is necessary to remove large glitches first to get close enough to the rms value that only represents the noise fluctuation.
We adapt $5~\sigma$ threshold to identify glitches, and the identified points are used to calculate the dead time fraction shown in Table~\ref{tab:1}.
It has been proven by simulations that the residuals of the cosmic rays at levels of $< 5~\sigma$ in the data do not affect the RMS noise or integration efficiency \cite{Catalano2016}.
The scheme is also cross-checked with a different method \cite{Baselmans2017} and the difference is small (less than 1~\% in the number of identified cosmic ray events).

Typical 10~s time traces of the MKID phase response for four arrays are shown in Fig.~\ref{fig:3}.
The identified glitches are highlighted as magenta points.
The ``KID ID" in the plots shows the order of MKIDs in readout frequency domain.
Although we have not carried out beam measurement to determine the position of each MKID in the chips, we can roughly identify the location from the resonance frequencies because of the encoding used in the design \cite{Baselmans2017}.
According to the encoding, KID ID from 300 to 600 roughly corresponds to the centre of the arrays, the lower and higher indices to the chip edge as indicated in Fig~\ref{fig:3}. 
For chip $a$, cosmic ray events are seen by a large fraction of the MKIDs.
On the other hand, the spread of glitches is suppressed for chip $b$, $c$, and $d$.

\begin{table*}
\begin{tabular}{l || c|c|c|c}
Array type & $a$ & $b$ & $c$ & $d$ \\
\hline \hline
Pixel yield [\%] & 85 & 85 & 82 & 83 \\
Dark NEP [$W/\sqrt{\rm{Hz}}$] ($\tau_{qp}$ [ms]) & ~$3 \times 10^{-19}$ (1.3)~ & ~$3 \times 10^{-19}$ (1.5)~ & ~$8 \times 10^{-19}$ (1.8)~ & ~$3 \times 10^{-19}$ (1.0)~ \\
Event rate per pixel [1/s]  & 0.69 & 0.089 & 0.10 & 0.028 \\
Affected pixels per event [\%] & 53 & 6.8 & 7.7 & 2.2 \\
Dead time fraction per pixel [\%] & 0.19 & 0.012 & 0.033 & 0.0046 \\
\hline
Estimated dead time at L2 [\%] & 23 & 1.4 & 4.0 & 0.55 \\
\end{tabular}
\caption{\label{tab:1}
Summary of the measurements.
The dead time fraction per pixel of chip $b$, $c$, and $d$ are reduced by factor of 16, 5.8 and 41 with respect to chip $a$, respectively.
}
\end{table*}

Table~\ref{tab:1} shows the summary of measurements.
About 80~\% out of 960 MKIDs are analyzed for each array, 20~\% are removed due to overlapping resonances or fit failures.
For each array we measure the dark Noise Equivalent Power (NEP) \cite{Baselmans2008} and $\tau_{qp}$ to demonstrate that the arrays are measured under the same conditions and have similar sensitivities.
Considering all glitches on the entire chip $a$, we measure an event rate of about 1.3~events/s ($\sim2.5$~events/min/cm$^2$), consistent with the expected muon rate. 
In order to make sure that events are really caused by cosmic rays, we count the number of events with a multiplicity more than 10, where ``multiplicity" is defined as the number of MKIDs that exceed the $5~\sigma$ threshold.
The main uncertainty of the event rate comes from solar activity.
For the experiments presented here, the change of solar activity was checked with neutron flux monitors \cite{NeutronMon} that correlate with the muon flux, and the variation is less than 5~\%.
Note that the neutron flux rate changes by 20-30~\% between minimum and maximum solar activity \cite{NeutronMon}.

The event rate per pixel in Table~\ref{tab:1} is calculated as: (Number of events in a pixel)~/~$T_{meas}$.
When evaluating the number of events, we apply a time threshold: When there is a glitch at a certain pixel at a certain time, the next $3~\tau_{qp}$ sequential time window is taken as 1 event if the response is larger than the $5~\sigma$ threshold.
The event rate per pixel is lower in all cases, showing that only a fraction of the pixels is affected for each cosmic ray interaction.
The fraction of affected pixels is also given in Table~\ref{tab:1}.
For instance, the event rate per pixel for chip $d$ is 0.028, so on average only $\sim0.028/1.3 = 2.2$~\% of the MKIDs are affected by a cosmic ray event whereas $\sim53$~\% of the pixels are affected in case of chip $a$.
The dead time fraction per pixel is equivalent to the fractional loss of integration time, and calculated as: (Number of glitch points in a pixel)~$\times~\Delta t_{read}~/~T_{meas}$.
The values shown in Table~\ref{tab:1} are the averaged value over all the analyzed MKIDs.

\begin{figure}
\includegraphics{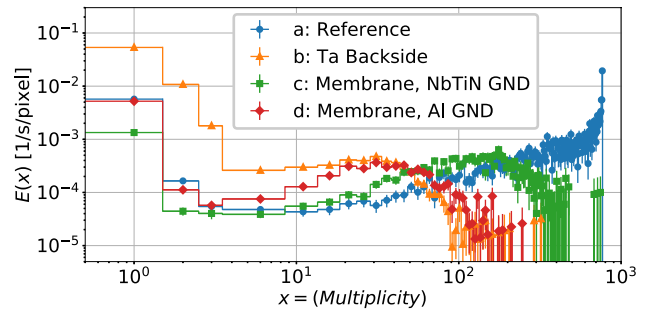}
\caption{\label{fig:4}
Measured probability density of event rate per pixel ($E(x)$) as a function of multiplicity ($x$).
Blue, orange, green, and red curves show chip $a$, $b$, $c$, and $d$, respectively.
The histograms are normalized so that the integral becomes event rate per pixel: $\int dx E(x) = (Event~rate~per~pixel)$.
The event rate per pixel for each arrays are shown in Table~\ref{tab:1}.
}
\end{figure}

Figure~\ref{fig:4} shows the event rate per pixel as a function of multiplicity.
It is clear that a large fraction of the events is caused by high multiplicity events in case of chip $a$ (blue).
In case of chip $b$ (orange), $c$ (green), and $d$ (red), the contribution from high multiplicity events is suppressed and it results in small values of event rate per pixel in Table~\ref{tab:1}.
Note that the probability density of single events ($x=1$) is quite high for all chips, but these events have a negligible contribution to the total dead time per chip.
For chip $a$, we have checked that the large value can be explained by statistics: the value is highly changed by changing the threshold of glitch detection.
For chip $b$, the large value is the result from scattered events as described in the caption of Fig.~\ref{fig:3}.
For array $c$ and $d$, the effect cannot be explained by statistics but may also involve the direct interaction between cosmic rays and the MKID aluminium, without phonon emission and propagation to the next pixel.

We can draw an important conclusion here.
If we do not adopt any measure to prevent phonon spread in the chip (as in chip $a$), the dead time fraction per pixel increases with the array size. 
On the other hand, a low $T_c$ phonon absorber and/or a membrane structure (chip $b$, $c$, and $d$) confine the spread of high energy phonons created by cosmic ray events, and the dead time fraction per pixel is independent of the array area: We can make larger-format arrays that do not suffer from significant data loss due to glitches created by cosmic rays.
There is another effect of the low $T_c$ layer: When comparing chip $a$ (blue, reference) with chip $b$ (orange, low $T_c$ layer), we observe a strong reduction in multiplicity.
The same effect is observed when comparing the membrane chips: The multiplicity of chip $c$ (green, NbTiN ground) is larger than $d$ (red, Al ground).

The data and techniques discussed in this letter are important not only for astronomical instruments, but also for quantum computing.
The key premises in error-corrected quantum computing are that errors can be sparse enough and therefore minimally correlated, and that logical error rates decrease exponentially with increasing system size \cite{Fowler2012}.
Cosmic ray events void both.
Figure~\ref{fig:3} (a) shows how cosmic ray events have generated quasiparticle excitations across the entire chip.
For superconducting qubits, the excess quasiparticles can cause deleterious state transitions \cite{Wenner2013, Rainis2012}, hence cosmic ray events may cause bursts of errors in time across the chip.
Moreover, the rate of cosmic ray events goes up with increasing chip size.
Completely blocking cosmic rays is impractical.
The decrease in event rate displayed in Fig.~\ref{fig:3} from using low gap materials and membranes, would carry over to a decrease in transition events associated with excess quasiparticles in superconducting quantum chips.
We therefore foresee that mitigating the effects of cosmic rays will become an important part in designing a quantum computer, and hope that the approach taken in this letter can act as guide.

Based on the obtained results, we can estimate the dead time fraction when operating the arrays in an L2 orbit by scaling the event rate on the entire chip (2.5~events/min/cm$^2$) to the event rate of 300~events/min/cm$^2$ that is reported by Planck \cite{Planck2014}. 
The calculated values are shown in Table~\ref{tab:1}.
We obtain a dead time fraction of 1.4~\% for a monolithic Si (chip $b$) with size of $\sim30$~cm$^2$ and an even smaller value of 0.55~\% for chip $d$.
These are much smaller than $\sim15$~\% data loss of Planck that uses Si dies with size of 0.4 to 0.8~cm$^2$ \cite{Planck2014, Catalano2014}.
We also simulate the energy deposition on a chip by cosmic rays at an L2 orbit assuming the energy spectrum given by Ref.~\onlinecite{Lotti2012} (see Fig.~\ref{fig:2}).
Although the primary component of cosmic rays at an L2 orbit is not muons but protons, the peak of the energy deposition from the simulation is almost the same.
This indicates that our simple scaling is a reasonable first order approximation.
However, the distribution of the deposited energy by protons is much broader with a high energy tail caused by low incident angle protons that travel longer through the Si wafer thereby depositing more energy.
In Fig.~\ref{fig:2} we also show a simulation of the spectral energy for a vertically placed chip on earth (orange line).
This increases the fraction of high deposited energies, at the cost of a much lower interaction rate.
For  a full understanding of cosmic ray interactions on large MKID arrays, such vertical lab measurements must be combined with dedicated beam line experiments to measure the effect of very high deposited energies which are possible in space.

In conclusion, we demonstrated a method to harden MKID arrays against cosmic ray events by adding a membrane structure and a layer of superconducting material with $T_c$ below or close to the $T_c$ of the Al of the MKIDs.
The idea to down-convert energy of phonons fully utilizes the advantage of MKIDs over bolometers: Phonons with energy lower than $2 \Delta_{Al}$ are invisible to MKIDs.
As a result, the dead time fraction per pixel caused by cosmic rays can be reduced up to a factor of 40 with respect to the reference array by a combination of a low $T_c$ layer and a membrane structure.
We have shown that these measures reduce the dead time fraction to less than 1~\% even at an L2 orbit.
The technique described in this paper is important not only for large-format MKID arrays for future astronomical instruments, but also for large qubit arrays for future quantum computers.

\begin{acknowledgments}
We would like to thank E. F. C. Driessen, A. Catalano, M. Calvo for useful discussion, and A. Tokiyasu for advices on GEANT4 simulation.
This work was supported by the ERC COG 648135 MOSAIC (J. J. A. B.), NWO Vidi Grant 639.042.423 (A. E.), and NWO Veni Grant 639.041.750 (P. J. de V.).
\end{acknowledgments}


\begin{thebibliography}{28}%
\makeatletter
\providecommand \@ifxundefined [1]{%
 \@ifx{#1\undefined}
}%
\providecommand \@ifnum [1]{%
 \ifnum #1\expandafter \@firstoftwo
 \else \expandafter \@secondoftwo
 \fi
}%
\providecommand \@ifx [1]{%
 \ifx #1\expandafter \@firstoftwo
 \else \expandafter \@secondoftwo
 \fi
}%
\providecommand \natexlab [1]{#1}%
\providecommand \enquote  [1]{``#1''}%
\providecommand \bibnamefont  [1]{#1}%
\providecommand \bibfnamefont [1]{#1}%
\providecommand \citenamefont [1]{#1}%
\providecommand \href@noop [0]{\@secondoftwo}%
\providecommand \href [0]{\begingroup \@sanitize@url \@href}%
\providecommand \@href[1]{\@@startlink{#1}\@@href}%
\providecommand \@@href[1]{\endgroup#1\@@endlink}%
\providecommand \@sanitize@url [0]{\catcode `\\12\catcode `\$12\catcode
  `\&12\catcode `\#12\catcode `\^12\catcode `\_12\catcode `\%12\relax}%
\providecommand \@@startlink[1]{}%
\providecommand \@@endlink[0]{}%
\providecommand \url  [0]{\begingroup\@sanitize@url \@url }%
\providecommand \@url [1]{\endgroup\@href {#1}{\urlprefix }}%
\providecommand \urlprefix  [0]{URL }%
\providecommand \Eprint [0]{\href }%
\providecommand \doibase [0]{http://dx.doi.org/}%
\providecommand \selectlanguage [0]{\@gobble}%
\providecommand \bibinfo  [0]{\@secondoftwo}%
\providecommand \bibfield  [0]{\@secondoftwo}%
\providecommand \translation [1]{[#1]}%
\providecommand \BibitemOpen [0]{}%
\providecommand \bibitemStop [0]{}%
\providecommand \bibitemNoStop [0]{.\EOS\space}%
\providecommand \EOS [0]{\spacefactor3000\relax}%
\providecommand \BibitemShut  [1]{\csname bibitem#1\endcsname}%
\let\auto@bib@innerbib\@empty
\bibitem [{\citenamefont {{Planck Collaboration}}(2014)}]{Planck2014}%
  \BibitemOpen
  \bibfield  {author} {\bibinfo {author} {\bibnamefont {{Planck
  Collaboration}}},\ }\href@noop {} {\bibfield  {journal} {\bibinfo  {journal}
  {A\&A}\ }\textbf {\bibinfo {volume} {571}},\ \bibinfo {pages} {A10} (\bibinfo
  {year} {2014})}\BibitemShut {NoStop}%
\bibitem [{\citenamefont {Day}\ \emph {et~al.}(2003)\citenamefont {Day},
  \citenamefont {LeDuc}, \citenamefont {Mazin}, \citenamefont {Vayonakis},\
  and\ \citenamefont {Zmuidzinas}}]{Day2003}%
  \BibitemOpen
  \bibfield  {author} {\bibinfo {author} {\bibfnamefont {P.~K.}\ \bibnamefont
  {Day}}, \bibinfo {author} {\bibfnamefont {H.~G.}\ \bibnamefont {LeDuc}},
  \bibinfo {author} {\bibfnamefont {B.~A.}\ \bibnamefont {Mazin}}, \bibinfo
  {author} {\bibfnamefont {A.}~\bibnamefont {Vayonakis}}, \ and\ \bibinfo
  {author} {\bibfnamefont {J.}~\bibnamefont {Zmuidzinas}},\ }\href@noop {}
  {\bibfield  {journal} {\bibinfo  {journal} {Nature}\ }\textbf {\bibinfo
  {volume} {425}},\ \bibinfo {pages} {817} (\bibinfo {year}
  {2003})}\BibitemShut {NoStop}%
\bibitem [{\citenamefont {Zmuidzinas}(2012)}]{Zmuidzinas2012}%
  \BibitemOpen
  \bibfield  {author} {\bibinfo {author} {\bibfnamefont {J.}~\bibnamefont
  {Zmuidzinas}},\ }\href@noop {} {\bibfield  {journal} {\bibinfo  {journal}
  {Annu. Rev. Cond. Mat. Phys.}\ }\textbf {\bibinfo {volume} {3}},\ \bibinfo
  {pages} {169} (\bibinfo {year} {2012})}\BibitemShut {NoStop}%
\bibitem [{\citenamefont {Irwin}(1995)}]{Irwin1995}%
  \BibitemOpen
  \bibfield  {author} {\bibinfo {author} {\bibfnamefont {K.~D.}\ \bibnamefont
  {Irwin}},\ }\href@noop {} {\bibfield  {journal} {\bibinfo  {journal} {Appl.
  Phys. Lett.}\ }\textbf {\bibinfo {volume} {66}},\ \bibinfo {pages} {1998}
  (\bibinfo {year} {1995})}\BibitemShut {NoStop}%
\bibitem [{\citenamefont {{Catalano}}\ \emph {et~al.}(2014)\citenamefont
  {{Catalano}}, \citenamefont {{Ade}}, \citenamefont {{Atik}}, \citenamefont
  {{Benoit}}, \citenamefont {{Br{\'e}ele}}, \citenamefont {{Bock}},
  \citenamefont {{Camus}}, \citenamefont {{Chabot}}, \citenamefont {{Charra}},
  \citenamefont {{Crill}}, \citenamefont {{Coron}}, \citenamefont {{Coulais}},
  \citenamefont {{D{\'e}sert}}, \citenamefont {{Fauvet}}, \citenamefont
  {{Giraud-H{\'e}raud}}, \citenamefont {{Guillaudin}}, \citenamefont
  {{Holmes}}, \citenamefont {{Jones}}, \citenamefont {{Lamarre}}, \citenamefont
  {{Mac{\'\i}as-P{\'e}rez}}, \citenamefont {{Martinez}}, \citenamefont
  {{Miniussi}}, \citenamefont {{Monfardini}}, \citenamefont {{Pajot}},
  \citenamefont {{Patanchon}}, \citenamefont {{Pelissier}}, \citenamefont
  {{Piat}}, \citenamefont {{Puget}}, \citenamefont {{Renault}}, \citenamefont
  {{Rosset}}, \citenamefont {{Santos}}, \citenamefont {{Sauv{\'e}}},
  \citenamefont {{Spencer}},\ and\ \citenamefont {{Sudiwala}}}]{Catalano2014}%
  \BibitemOpen
  \bibfield  {author} {\bibinfo {author} {\bibfnamefont {A.}~\bibnamefont
  {{Catalano}}}, \bibinfo {author} {\bibfnamefont {P.}~\bibnamefont {{Ade}}},
  \bibinfo {author} {\bibfnamefont {Y.}~\bibnamefont {{Atik}}}, \bibinfo
  {author} {\bibfnamefont {A.}~\bibnamefont {{Benoit}}}, \bibinfo {author}
  {\bibfnamefont {E.}~\bibnamefont {{Br{\'e}ele}}}, \bibinfo {author}
  {\bibfnamefont {J.~J.}\ \bibnamefont {{Bock}}}, \bibinfo {author}
  {\bibfnamefont {P.}~\bibnamefont {{Camus}}}, \bibinfo {author} {\bibfnamefont
  {M.}~\bibnamefont {{Chabot}}}, \bibinfo {author} {\bibfnamefont
  {M.}~\bibnamefont {{Charra}}}, \bibinfo {author} {\bibfnamefont {B.~P.}\
  \bibnamefont {{Crill}}}, \bibinfo {author} {\bibfnamefont {N.}~\bibnamefont
  {{Coron}}}, \bibinfo {author} {\bibfnamefont {A.}~\bibnamefont {{Coulais}}},
  \bibinfo {author} {\bibfnamefont {F.~X.}\ \bibnamefont {{D{\'e}sert}}},
  \bibinfo {author} {\bibfnamefont {L.}~\bibnamefont {{Fauvet}}}, \bibinfo
  {author} {\bibfnamefont {Y.}~\bibnamefont {{Giraud-H{\'e}raud}}}, \bibinfo
  {author} {\bibfnamefont {O.}~\bibnamefont {{Guillaudin}}}, \bibinfo {author}
  {\bibfnamefont {W.}~\bibnamefont {{Holmes}}}, \bibinfo {author}
  {\bibfnamefont {W.~C.}\ \bibnamefont {{Jones}}}, \bibinfo {author}
  {\bibfnamefont {J.~M.}\ \bibnamefont {{Lamarre}}}, \bibinfo {author}
  {\bibfnamefont {J.}~\bibnamefont {{Mac{\'\i}as-P{\'e}rez}}}, \bibinfo
  {author} {\bibfnamefont {M.}~\bibnamefont {{Martinez}}}, \bibinfo {author}
  {\bibfnamefont {A.}~\bibnamefont {{Miniussi}}}, \bibinfo {author}
  {\bibfnamefont {A.}~\bibnamefont {{Monfardini}}}, \bibinfo {author}
  {\bibfnamefont {F.}~\bibnamefont {{Pajot}}}, \bibinfo {author} {\bibfnamefont
  {G.}~\bibnamefont {{Patanchon}}}, \bibinfo {author} {\bibfnamefont
  {A.}~\bibnamefont {{Pelissier}}}, \bibinfo {author} {\bibfnamefont
  {M.}~\bibnamefont {{Piat}}}, \bibinfo {author} {\bibfnamefont {J.~L.}\
  \bibnamefont {{Puget}}}, \bibinfo {author} {\bibfnamefont {C.}~\bibnamefont
  {{Renault}}}, \bibinfo {author} {\bibfnamefont {C.}~\bibnamefont {{Rosset}}},
  \bibinfo {author} {\bibfnamefont {D.}~\bibnamefont {{Santos}}}, \bibinfo
  {author} {\bibfnamefont {A.}~\bibnamefont {{Sauv{\'e}}}}, \bibinfo {author}
  {\bibfnamefont {L.~D.}\ \bibnamefont {{Spencer}}}, \ and\ \bibinfo {author}
  {\bibfnamefont {R.}~\bibnamefont {{Sudiwala}}},\ }\href@noop {} {\bibfield
  {journal} {\bibinfo  {journal} {A\&A}\ }\textbf {\bibinfo {volume} {569}},\
  \bibinfo {pages} {A88} (\bibinfo {year} {2014})}\BibitemShut {NoStop}%
\bibitem [{\citenamefont {{Catalano}}\ \emph {et~al.}(2016)\citenamefont
  {{Catalano}}, \citenamefont {{Benoit}}, \citenamefont {{Bourrion}},
  \citenamefont {{Calvo}}, \citenamefont {{Coiffard}}, \citenamefont
  {{D'Addabbo}}, \citenamefont {{Goupy}}, \citenamefont {{LeSueur}},
  \citenamefont {{Mac{\'\i}as-P{\'e}rez}},\ and\ \citenamefont
  {{Monfardini}}}]{Catalano2016}%
  \BibitemOpen
  \bibfield  {author} {\bibinfo {author} {\bibfnamefont {A.}~\bibnamefont
  {{Catalano}}}, \bibinfo {author} {\bibfnamefont {A.}~\bibnamefont
  {{Benoit}}}, \bibinfo {author} {\bibfnamefont {O.}~\bibnamefont
  {{Bourrion}}}, \bibinfo {author} {\bibfnamefont {M.}~\bibnamefont {{Calvo}}},
  \bibinfo {author} {\bibfnamefont {G.}~\bibnamefont {{Coiffard}}}, \bibinfo
  {author} {\bibfnamefont {A.}~\bibnamefont {{D'Addabbo}}}, \bibinfo {author}
  {\bibfnamefont {J.}~\bibnamefont {{Goupy}}}, \bibinfo {author} {\bibfnamefont
  {H.}~\bibnamefont {{LeSueur}}}, \bibinfo {author} {\bibfnamefont
  {J.}~\bibnamefont {{Mac{\'\i}as-P{\'e}rez}}}, \ and\ \bibinfo {author}
  {\bibfnamefont {A.}~\bibnamefont {{Monfardini}}},\ }\href@noop {} {\bibfield
  {journal} {\bibinfo  {journal} {A\&A}\ }\textbf {\bibinfo {volume} {592}}
  (\bibinfo {year} {2016})}\BibitemShut {NoStop}%
\bibitem [{\citenamefont {{de Visser}}\ \emph {et~al.}(2014)\citenamefont {{de
  Visser}}, \citenamefont {{Baselmans}}, \citenamefont {{Bueno}}, \citenamefont
  {{Llombart}},\ and\ \citenamefont {{Klapwijk}}}]{deVisser2014}%
  \BibitemOpen
  \bibfield  {author} {\bibinfo {author} {\bibfnamefont {P.~J.}\ \bibnamefont
  {{de Visser}}}, \bibinfo {author} {\bibfnamefont {J.~J.~A.}\ \bibnamefont
  {{Baselmans}}}, \bibinfo {author} {\bibfnamefont {J.}~\bibnamefont
  {{Bueno}}}, \bibinfo {author} {\bibfnamefont {N.}~\bibnamefont {{Llombart}}},
  \ and\ \bibinfo {author} {\bibfnamefont {T.~M.}\ \bibnamefont {{Klapwijk}}},\
  }\href@noop {} {\bibfield  {journal} {\bibinfo  {journal} {Nature
  Communications}\ }\textbf {\bibinfo {volume} {5}},\ \bibinfo {pages} {3130}
  (\bibinfo {year} {2014})}\BibitemShut {NoStop}%
\bibitem [{\citenamefont {Janssen}\ \emph {et~al.}(2013)\citenamefont
  {Janssen}, \citenamefont {Baselmans}, \citenamefont {Endo}, \citenamefont
  {Ferrari}, \citenamefont {Yates}, \citenamefont {Baryshev},\ and\
  \citenamefont {Klapwijk}}]{Janssen2013}%
  \BibitemOpen
  \bibfield  {author} {\bibinfo {author} {\bibfnamefont {R.~M.~J.}\
  \bibnamefont {Janssen}}, \bibinfo {author} {\bibfnamefont {J.~J.~A.}\
  \bibnamefont {Baselmans}}, \bibinfo {author} {\bibfnamefont {A.}~\bibnamefont
  {Endo}}, \bibinfo {author} {\bibfnamefont {L.}~\bibnamefont {Ferrari}},
  \bibinfo {author} {\bibfnamefont {S.~J.~C.}\ \bibnamefont {Yates}}, \bibinfo
  {author} {\bibfnamefont {A.~M.}\ \bibnamefont {Baryshev}}, \ and\ \bibinfo
  {author} {\bibfnamefont {T.~M.}\ \bibnamefont {Klapwijk}},\ }\href@noop {}
  {\bibfield  {journal} {\bibinfo  {journal} {Appl. Phys. Lett.}\ }\textbf
  {\bibinfo {volume} {103}},\ \bibinfo {pages} {203503} (\bibinfo {year}
  {2013})}\BibitemShut {NoStop}%
\bibitem [{\citenamefont {{Baselmans}}\ \emph {et~al.}(2017)\citenamefont
  {{Baselmans}}, \citenamefont {{Bueno}}, \citenamefont {{Yates}},
  \citenamefont {{Yurduseven}}, \citenamefont {{Llombart}}, \citenamefont
  {{Karatsu}}, \citenamefont {{Baryshev}}, \citenamefont {{Ferrari}},
  \citenamefont {{Endo}}, \citenamefont {{Thoen}}, \citenamefont {{de Visser}},
  \citenamefont {{Janssen}}, \citenamefont {{Murugesan}}, \citenamefont
  {{Driessen}}, \citenamefont {{Coiffard}}, \citenamefont {{Martin-Pintado}},
  \citenamefont {{Hargrave}},\ and\ \citenamefont {{Griffin}}}]{Baselmans2017}%
  \BibitemOpen
  \bibfield  {author} {\bibinfo {author} {\bibfnamefont {J.~J.~A.}\
  \bibnamefont {{Baselmans}}}, \bibinfo {author} {\bibfnamefont
  {J.}~\bibnamefont {{Bueno}}}, \bibinfo {author} {\bibfnamefont {S.~J.~C.}\
  \bibnamefont {{Yates}}}, \bibinfo {author} {\bibfnamefont {O.}~\bibnamefont
  {{Yurduseven}}}, \bibinfo {author} {\bibfnamefont {N.}~\bibnamefont
  {{Llombart}}}, \bibinfo {author} {\bibfnamefont {K.}~\bibnamefont
  {{Karatsu}}}, \bibinfo {author} {\bibfnamefont {A.~M.}\ \bibnamefont
  {{Baryshev}}}, \bibinfo {author} {\bibfnamefont {L.}~\bibnamefont
  {{Ferrari}}}, \bibinfo {author} {\bibfnamefont {A.}~\bibnamefont {{Endo}}},
  \bibinfo {author} {\bibfnamefont {D.~J.}\ \bibnamefont {{Thoen}}}, \bibinfo
  {author} {\bibfnamefont {P.~J.}\ \bibnamefont {{de Visser}}}, \bibinfo
  {author} {\bibfnamefont {R.~M.~J.}\ \bibnamefont {{Janssen}}}, \bibinfo
  {author} {\bibfnamefont {V.}~\bibnamefont {{Murugesan}}}, \bibinfo {author}
  {\bibfnamefont {E.~F.~C.}\ \bibnamefont {{Driessen}}}, \bibinfo {author}
  {\bibfnamefont {G.}~\bibnamefont {{Coiffard}}}, \bibinfo {author}
  {\bibfnamefont {J.}~\bibnamefont {{Martin-Pintado}}}, \bibinfo {author}
  {\bibfnamefont {P.}~\bibnamefont {{Hargrave}}}, \ and\ \bibinfo {author}
  {\bibfnamefont {M.}~\bibnamefont {{Griffin}}},\ }\href@noop {} {\bibfield
  {journal} {\bibinfo  {journal} {A\&A}\ }\textbf {\bibinfo {volume} {601}}
  (\bibinfo {year} {2017})}\BibitemShut {NoStop}%
\bibitem [{\citenamefont {Yates}\ \emph {et~al.}(2017)\citenamefont {Yates},
  \citenamefont {Baryshev}, \citenamefont {Yurduseven}, \citenamefont {Bueno},
  \citenamefont {Davis}, \citenamefont {Ferrari}, \citenamefont {Jellema},
  \citenamefont {Llombart}, \citenamefont {Murugesan}, \citenamefont {Thoen},\
  and\ \citenamefont {Baselmans}}]{Yates2017}%
  \BibitemOpen
  \bibfield  {author} {\bibinfo {author} {\bibfnamefont {S.~J.~C.}\
  \bibnamefont {Yates}}, \bibinfo {author} {\bibfnamefont {A.~M.}\ \bibnamefont
  {Baryshev}}, \bibinfo {author} {\bibfnamefont {O.}~\bibnamefont
  {Yurduseven}}, \bibinfo {author} {\bibfnamefont {J.}~\bibnamefont {Bueno}},
  \bibinfo {author} {\bibfnamefont {K.~K.}\ \bibnamefont {Davis}}, \bibinfo
  {author} {\bibfnamefont {L.}~\bibnamefont {Ferrari}}, \bibinfo {author}
  {\bibfnamefont {W.}~\bibnamefont {Jellema}}, \bibinfo {author} {\bibfnamefont
  {N.}~\bibnamefont {Llombart}}, \bibinfo {author} {\bibfnamefont
  {V.}~\bibnamefont {Murugesan}}, \bibinfo {author} {\bibfnamefont {D.~J.}\
  \bibnamefont {Thoen}}, \ and\ \bibinfo {author} {\bibfnamefont {J.~J.~A.}\
  \bibnamefont {Baselmans}},\ }\href@noop {} {\bibfield  {journal} {\bibinfo
  {journal} {IEEE Trans. on Terahertz Science and Technology}\ }\textbf
  {\bibinfo {volume} {7}},\ \bibinfo {pages} {789} (\bibinfo {year}
  {2017})}\BibitemShut {NoStop}%
\bibitem [{\citenamefont {{H\"ahnle}}\ \emph {et~al.}()\citenamefont
  {{H\"ahnle}}, \citenamefont {{Yurduseven}}, \citenamefont {{van Berkel}},
  \citenamefont {{Llombart}}, \citenamefont {{Bueno}}, \citenamefont {{Yates}},
  \citenamefont {{Murugesan}}, \citenamefont {{Thoen}}, \citenamefont
  {{Neto}},\ and\ \citenamefont {{Baselmans}}}]{Hahnle2018}%
  \BibitemOpen
  \bibfield  {author} {\bibinfo {author} {\bibfnamefont {S.}~\bibnamefont
  {{H\"ahnle}}}, \bibinfo {author} {\bibfnamefont {O.}~\bibnamefont
  {{Yurduseven}}}, \bibinfo {author} {\bibfnamefont {S.~L.}\ \bibnamefont {{van
  Berkel}}}, \bibinfo {author} {\bibfnamefont {N.}~\bibnamefont {{Llombart}}},
  \bibinfo {author} {\bibfnamefont {J.}~\bibnamefont {{Bueno}}}, \bibinfo
  {author} {\bibfnamefont {S.~J.~C.}\ \bibnamefont {{Yates}}}, \bibinfo
  {author} {\bibfnamefont {V.}~\bibnamefont {{Murugesan}}}, \bibinfo {author}
  {\bibfnamefont {D.~J.}\ \bibnamefont {{Thoen}}}, \bibinfo {author}
  {\bibfnamefont {A.}~\bibnamefont {{Neto}}}, \ and\ \bibinfo {author}
  {\bibfnamefont {J.~J.~A.}\ \bibnamefont {{Baselmans}}},\ }\href@noop {}
  {}\bibinfo {note} {In preparation}\BibitemShut {NoStop}%
\bibitem [{\citenamefont {Bueno}\ \emph {et~al.}(2017)\citenamefont {Bueno},
  \citenamefont {Yurduseven}, \citenamefont {Yates}, \citenamefont {Llombart},
  \citenamefont {Murugesan}, \citenamefont {Thoen}, \citenamefont {Baryshev},
  \citenamefont {Neto},\ and\ \citenamefont {Baselmans}}]{Bueno2017a}%
  \BibitemOpen
  \bibfield  {author} {\bibinfo {author} {\bibfnamefont {J.}~\bibnamefont
  {Bueno}}, \bibinfo {author} {\bibfnamefont {O.}~\bibnamefont {Yurduseven}},
  \bibinfo {author} {\bibfnamefont {S.~J.~C.}\ \bibnamefont {Yates}}, \bibinfo
  {author} {\bibfnamefont {N.}~\bibnamefont {Llombart}}, \bibinfo {author}
  {\bibfnamefont {V.}~\bibnamefont {Murugesan}}, \bibinfo {author}
  {\bibfnamefont {D.~J.}\ \bibnamefont {Thoen}}, \bibinfo {author}
  {\bibfnamefont {A.~M.}\ \bibnamefont {Baryshev}}, \bibinfo {author}
  {\bibfnamefont {A.}~\bibnamefont {Neto}}, \ and\ \bibinfo {author}
  {\bibfnamefont {J.~J.~A.}\ \bibnamefont {Baselmans}},\ }\href@noop {}
  {\bibfield  {journal} {\bibinfo  {journal} {Appl. Phys. Lett.}\ }\textbf
  {\bibinfo {volume} {110}},\ \bibinfo {pages} {233503} (\bibinfo {year}
  {2017})}\BibitemShut {NoStop}%
\bibitem [{\citenamefont {Bueno}\ \emph {et~al.}(2018)\citenamefont {Bueno},
  \citenamefont {Murugesan}, \citenamefont {Karatsu}, \citenamefont {Thoen},\
  and\ \citenamefont {Baselmans}}]{Bueno2017b}%
  \BibitemOpen
  \bibfield  {author} {\bibinfo {author} {\bibfnamefont {J.}~\bibnamefont
  {Bueno}}, \bibinfo {author} {\bibfnamefont {V.}~\bibnamefont {Murugesan}},
  \bibinfo {author} {\bibfnamefont {K.}~\bibnamefont {Karatsu}}, \bibinfo
  {author} {\bibfnamefont {D.~J.}\ \bibnamefont {Thoen}}, \ and\ \bibinfo
  {author} {\bibfnamefont {J.~J.~A.}\ \bibnamefont {Baselmans}},\ }\href
  {\doibase 10.1007/s10909-018-1962-8} {\bibfield  {journal} {\bibinfo
  {journal} {J. Low Temp. Phys.}\ } (\bibinfo {year} {2018}),\
  10.1007/s10909-018-1962-8}\BibitemShut {NoStop}%
\bibitem [{\citenamefont {Monfardini}\ \emph {et~al.}(2016)\citenamefont
  {Monfardini}, \citenamefont {Baselmans}, \citenamefont {Benoit},
  \citenamefont {Bideaud}, \citenamefont {Bourrion}, \citenamefont {Catalano},
  \citenamefont {Calvo}, \citenamefont {D'Addabbo}, \citenamefont {Doyle},
  \citenamefont {Goupy}, \citenamefont {Sueur},\ and\ \citenamefont
  {Macias-Perez}}]{Monfardini2016}%
  \BibitemOpen
  \bibfield  {author} {\bibinfo {author} {\bibfnamefont {A.}~\bibnamefont
  {Monfardini}}, \bibinfo {author} {\bibfnamefont {J.~J.~A.}\ \bibnamefont
  {Baselmans}}, \bibinfo {author} {\bibfnamefont {A.}~\bibnamefont {Benoit}},
  \bibinfo {author} {\bibfnamefont {A.}~\bibnamefont {Bideaud}}, \bibinfo
  {author} {\bibfnamefont {O.}~\bibnamefont {Bourrion}}, \bibinfo {author}
  {\bibfnamefont {A.}~\bibnamefont {Catalano}}, \bibinfo {author}
  {\bibfnamefont {M.}~\bibnamefont {Calvo}}, \bibinfo {author} {\bibfnamefont
  {A.}~\bibnamefont {D'Addabbo}}, \bibinfo {author} {\bibfnamefont
  {S.}~\bibnamefont {Doyle}}, \bibinfo {author} {\bibfnamefont
  {J.}~\bibnamefont {Goupy}}, \bibinfo {author} {\bibfnamefont {H.~L.}\
  \bibnamefont {Sueur}}, \ and\ \bibinfo {author} {\bibfnamefont
  {J.}~\bibnamefont {Macias-Perez}},\ }\href@noop {} {\bibfield  {journal}
  {\bibinfo  {journal} {Proc. SPIE}\ }\textbf {\bibinfo {volume} {9914}},\
  \bibinfo {pages} {99140} (\bibinfo {year} {2016})}\BibitemShut {NoStop}%
\bibitem [{\citenamefont {Kozorezov}\ \emph {et~al.}(2000)\citenamefont
  {Kozorezov}, \citenamefont {Volkov}, \citenamefont {Wigmore}, \citenamefont
  {Peacock}, \citenamefont {Poelaert},\ and\ \citenamefont {den
  Hartog}}]{Kozorezov2000}%
  \BibitemOpen
  \bibfield  {author} {\bibinfo {author} {\bibfnamefont {A.~G.}\ \bibnamefont
  {Kozorezov}}, \bibinfo {author} {\bibfnamefont {A.~F.}\ \bibnamefont
  {Volkov}}, \bibinfo {author} {\bibfnamefont {J.~K.}\ \bibnamefont {Wigmore}},
  \bibinfo {author} {\bibfnamefont {A.}~\bibnamefont {Peacock}}, \bibinfo
  {author} {\bibfnamefont {A.}~\bibnamefont {Poelaert}}, \ and\ \bibinfo
  {author} {\bibfnamefont {R.}~\bibnamefont {den Hartog}},\ }\href@noop {}
  {\bibfield  {journal} {\bibinfo  {journal} {Phys. Rev. B}\ }\textbf {\bibinfo
  {volume} {61}},\ \bibinfo {pages} {11807} (\bibinfo {year}
  {2000})}\BibitemShut {NoStop}%
\bibitem [{\citenamefont {Goldie}\ and\ \citenamefont
  {Withington}(2013)}]{Goldie2013}%
  \BibitemOpen
  \bibfield  {author} {\bibinfo {author} {\bibfnamefont {D.~J.}\ \bibnamefont
  {Goldie}}\ and\ \bibinfo {author} {\bibfnamefont {S.}~\bibnamefont
  {Withington}},\ }\href@noop {} {\bibfield  {journal} {\bibinfo  {journal}
  {Supercond. Sci. Technol.}\ }\textbf {\bibinfo {volume} {26}},\ \bibinfo
  {pages} {015004} (\bibinfo {year} {2013})}\BibitemShut {NoStop}%
\bibitem [{\citenamefont {Holmes}\ \emph {et~al.}(2008)\citenamefont {Holmes},
  \citenamefont {Bock}, \citenamefont {Crill}, \citenamefont {Koch},
  \citenamefont {Jones}, \citenamefont {Lange},\ and\ \citenamefont
  {Paine}}]{Holmes2008}%
  \BibitemOpen
  \bibfield  {author} {\bibinfo {author} {\bibfnamefont {W.~A.}\ \bibnamefont
  {Holmes}}, \bibinfo {author} {\bibfnamefont {J.~J.}\ \bibnamefont {Bock}},
  \bibinfo {author} {\bibfnamefont {B.~P.}\ \bibnamefont {Crill}}, \bibinfo
  {author} {\bibfnamefont {T.~C.}\ \bibnamefont {Koch}}, \bibinfo {author}
  {\bibfnamefont {W.~C.}\ \bibnamefont {Jones}}, \bibinfo {author}
  {\bibfnamefont {A.~E.}\ \bibnamefont {Lange}}, \ and\ \bibinfo {author}
  {\bibfnamefont {C.~G.}\ \bibnamefont {Paine}},\ }\href@noop {} {\bibfield
  {journal} {\bibinfo  {journal} {Appl. Opt.}\ }\textbf {\bibinfo {volume}
  {47}},\ \bibinfo {pages} {5996} (\bibinfo {year} {2008})}\BibitemShut
  {NoStop}%
\bibitem [{\citenamefont {Baselmans}\ \emph {et~al.}(2012)\citenamefont
  {Baselmans}, \citenamefont {Yates}, \citenamefont {Diener},\ and\
  \citenamefont {de~Visser}}]{Baselmans2012}%
  \BibitemOpen
  \bibfield  {author} {\bibinfo {author} {\bibfnamefont {J.~J.~A.}\
  \bibnamefont {Baselmans}}, \bibinfo {author} {\bibfnamefont {S.~J.~C.}\
  \bibnamefont {Yates}}, \bibinfo {author} {\bibfnamefont {P.}~\bibnamefont
  {Diener}}, \ and\ \bibinfo {author} {\bibfnamefont {P.~J.}\ \bibnamefont
  {de~Visser}},\ }\href@noop {} {\bibfield  {journal} {\bibinfo  {journal} {J.
  Low Temp. Phys.}\ }\textbf {\bibinfo {volume} {167}},\ \bibinfo {pages} {360}
  (\bibinfo {year} {2012})}\BibitemShut {NoStop}%
\bibitem [{Note1()}]{Note1}%
  \BibitemOpen
  \bibinfo {note} {Chip $a$ and $b$ has a gold layer at edges to have a good
  thermal contact to the gold-plated sample box. The layer is intended to
  better thermalize the chip under sub-mm illumination \cite
  {Yates2017}.}\BibitemShut {Stop}%
\bibitem [{\citenamefont {van Rantwijk}\ \emph {et~al.}(2016)\citenamefont {van
  Rantwijk}, \citenamefont {Grim}, \citenamefont {van Loon}, \citenamefont
  {Yates}, \citenamefont {Baryshev},\ and\ \citenamefont
  {Baselmans}}]{Rantwijk2016}%
  \BibitemOpen
  \bibfield  {author} {\bibinfo {author} {\bibfnamefont {J.}~\bibnamefont {van
  Rantwijk}}, \bibinfo {author} {\bibfnamefont {M.}~\bibnamefont {Grim}},
  \bibinfo {author} {\bibfnamefont {D.}~\bibnamefont {van Loon}}, \bibinfo
  {author} {\bibfnamefont {S.~J.~C.}\ \bibnamefont {Yates}}, \bibinfo {author}
  {\bibfnamefont {A.~M.}\ \bibnamefont {Baryshev}}, \ and\ \bibinfo {author}
  {\bibfnamefont {J.~J.~A.}\ \bibnamefont {Baselmans}},\ }\href@noop {}
  {\bibfield  {journal} {\bibinfo  {journal} {IEEE Trans. on Microwave Theory
  and Techniques}\ }\textbf {\bibinfo {volume} {64}},\ \bibinfo {pages} {1876}
  (\bibinfo {year} {2016})}\BibitemShut {NoStop}%
\bibitem [{\citenamefont {{GEANT4 Collaboration}}(2003)}]{GEANT2003}%
  \BibitemOpen
  \bibfield  {author} {\bibinfo {author} {\bibnamefont {{GEANT4
  Collaboration}}},\ }\href@noop {} {\bibfield  {journal} {\bibinfo  {journal}
  {Nucl. Instrum. Meth.}\ }\textbf {\bibinfo {volume} {A506}},\ \bibinfo
  {pages} {250} (\bibinfo {year} {2003})}\BibitemShut {NoStop}%
\bibitem [{CRY()}]{CRY}%
  \BibitemOpen
  \href@noop {} {}\bibinfo {howpublished}
  {https://nuclear.llnl.gov/simulation/}\BibitemShut {NoStop}%
\bibitem [{\citenamefont {Baselmans}\ \emph {et~al.}(2008)\citenamefont
  {Baselmans}, \citenamefont {Yates}, \citenamefont {Barends}, \citenamefont
  {Lankwarden}, \citenamefont {Gao}, \citenamefont {Hoevers},\ and\
  \citenamefont {Klapwijk}}]{Baselmans2008}%
  \BibitemOpen
  \bibfield  {author} {\bibinfo {author} {\bibfnamefont {J.~J.~A.}\
  \bibnamefont {Baselmans}}, \bibinfo {author} {\bibfnamefont {S.~J.~C.}\
  \bibnamefont {Yates}}, \bibinfo {author} {\bibfnamefont {R.}~\bibnamefont
  {Barends}}, \bibinfo {author} {\bibfnamefont {Y.~J.~Y.}\ \bibnamefont
  {Lankwarden}}, \bibinfo {author} {\bibfnamefont {J.~R.}\ \bibnamefont {Gao}},
  \bibinfo {author} {\bibfnamefont {H.}~\bibnamefont {Hoevers}}, \ and\
  \bibinfo {author} {\bibfnamefont {T.~M.}\ \bibnamefont {Klapwijk}},\
  }\href@noop {} {\bibfield  {journal} {\bibinfo  {journal} {J. Low Temp.
  Phys.}\ }\textbf {\bibinfo {volume} {151}},\ \bibinfo {pages} {524} (\bibinfo
  {year} {2008})}\BibitemShut {NoStop}%
\bibitem [{Neu()}]{NeutronMon}%
  \BibitemOpen
  \href@noop {} {}\bibinfo {howpublished}
  {https://neutronm.bartol.udel.edu/}\BibitemShut {NoStop}%
\bibitem [{\citenamefont {Fowler}\ \emph {et~al.}(2012)\citenamefont {Fowler},
  \citenamefont {Mariantoni}, \citenamefont {Martinis},\ and\ \citenamefont
  {Cleland}}]{Fowler2012}%
  \BibitemOpen
  \bibfield  {author} {\bibinfo {author} {\bibfnamefont {A.~G.}\ \bibnamefont
  {Fowler}}, \bibinfo {author} {\bibfnamefont {M.}~\bibnamefont {Mariantoni}},
  \bibinfo {author} {\bibfnamefont {J.~M.}\ \bibnamefont {Martinis}}, \ and\
  \bibinfo {author} {\bibfnamefont {A.~N.}\ \bibnamefont {Cleland}},\
  }\href@noop {} {\bibfield  {journal} {\bibinfo  {journal} {Phys. Rev. A}\
  }\textbf {\bibinfo {volume} {86}},\ \bibinfo {pages} {032324} (\bibinfo
  {year} {2012})}\BibitemShut {NoStop}%
\bibitem [{\citenamefont {Wenner}\ \emph {et~al.}(2013)\citenamefont {Wenner},
  \citenamefont {Yin}, \citenamefont {Lucero}, \citenamefont {Barends},
  \citenamefont {Chen}, \citenamefont {Chiaro}, \citenamefont {Kelly},
  \citenamefont {Lenander}, \citenamefont {Mariantoni}, \citenamefont
  {Megrant}, \citenamefont {Neill}, \citenamefont {O'Malley}, \citenamefont
  {Sank}, \citenamefont {Vainsencher}, \citenamefont {Wang}, \citenamefont
  {White}, \citenamefont {Cleland},\ and\ \citenamefont
  {Martinis}}]{Wenner2013}%
  \BibitemOpen
  \bibfield  {author} {\bibinfo {author} {\bibfnamefont {J.}~\bibnamefont
  {Wenner}}, \bibinfo {author} {\bibfnamefont {Y.}~\bibnamefont {Yin}},
  \bibinfo {author} {\bibfnamefont {E.}~\bibnamefont {Lucero}}, \bibinfo
  {author} {\bibfnamefont {R.}~\bibnamefont {Barends}}, \bibinfo {author}
  {\bibfnamefont {Y.}~\bibnamefont {Chen}}, \bibinfo {author} {\bibfnamefont
  {B.}~\bibnamefont {Chiaro}}, \bibinfo {author} {\bibfnamefont
  {J.}~\bibnamefont {Kelly}}, \bibinfo {author} {\bibfnamefont
  {M.}~\bibnamefont {Lenander}}, \bibinfo {author} {\bibfnamefont
  {M.}~\bibnamefont {Mariantoni}}, \bibinfo {author} {\bibfnamefont
  {A.}~\bibnamefont {Megrant}}, \bibinfo {author} {\bibfnamefont
  {C.}~\bibnamefont {Neill}}, \bibinfo {author} {\bibfnamefont {P.~J.~J.}\
  \bibnamefont {O'Malley}}, \bibinfo {author} {\bibfnamefont {D.}~\bibnamefont
  {Sank}}, \bibinfo {author} {\bibfnamefont {A.}~\bibnamefont {Vainsencher}},
  \bibinfo {author} {\bibfnamefont {H.}~\bibnamefont {Wang}}, \bibinfo {author}
  {\bibfnamefont {T.~C.}\ \bibnamefont {White}}, \bibinfo {author}
  {\bibfnamefont {A.~N.}\ \bibnamefont {Cleland}}, \ and\ \bibinfo {author}
  {\bibfnamefont {J.~M.}\ \bibnamefont {Martinis}},\ }\href@noop {} {\bibfield
  {journal} {\bibinfo  {journal} {Phys. Rev. Lett.}\ }\textbf {\bibinfo
  {volume} {110}},\ \bibinfo {pages} {150502} (\bibinfo {year}
  {2013})}\BibitemShut {NoStop}%
\bibitem [{\citenamefont {Rainis}\ and\ \citenamefont
  {Loss}(2012)}]{Rainis2012}%
  \BibitemOpen
  \bibfield  {author} {\bibinfo {author} {\bibfnamefont {D.}~\bibnamefont
  {Rainis}}\ and\ \bibinfo {author} {\bibfnamefont {D.}~\bibnamefont {Loss}},\
  }\href@noop {} {\bibfield  {journal} {\bibinfo  {journal} {Phys. Rev. B}\
  }\textbf {\bibinfo {volume} {85}},\ \bibinfo {pages} {174533} (\bibinfo
  {year} {2012})}\BibitemShut {NoStop}%
\bibitem [{\citenamefont {Lotti}\ \emph {et~al.}(2012)\citenamefont {Lotti},
  \citenamefont {Perinati}, \citenamefont {Natalucci}, \citenamefont {Piro},
  \citenamefont {Mineo}, \citenamefont {Colasanti},\ and\ \citenamefont
  {Macculi}}]{Lotti2012}%
  \BibitemOpen
  \bibfield  {author} {\bibinfo {author} {\bibfnamefont {S.}~\bibnamefont
  {Lotti}}, \bibinfo {author} {\bibfnamefont {E.}~\bibnamefont {Perinati}},
  \bibinfo {author} {\bibfnamefont {L.}~\bibnamefont {Natalucci}}, \bibinfo
  {author} {\bibfnamefont {L.}~\bibnamefont {Piro}}, \bibinfo {author}
  {\bibfnamefont {T.}~\bibnamefont {Mineo}}, \bibinfo {author} {\bibfnamefont
  {L.}~\bibnamefont {Colasanti}}, \ and\ \bibinfo {author} {\bibfnamefont
  {C.}~\bibnamefont {Macculi}},\ }\href@noop {} {\bibfield  {journal} {\bibinfo
   {journal} {Nucl. Instrum. Meth.}\ }\textbf {\bibinfo {volume} {A686}},\
  \bibinfo {pages} {31 } (\bibinfo {year} {2012})}\BibitemShut {NoStop}%
\end{thebibliography}
\providecommand{\noopsort}[1]{}\providecommand{\singleletter}[1]{#1}%

\end{document}